\documentstyle[preprint,aps]{revtex}
\begin{document}

\draft
\title{RENORMALIZATION GROUP APPROACH TO THE COULOMB PSEUDOPOTENTIAL FOR
${\rm C}_{60}$}

\author{Nikos Berdenis and Ganpathy Murthy}

\address{Department of Physics, Boston University}

\maketitle

\begin{abstract}
A numerical renormalization group technique recently developed by one of
us is used to analyse the Coulomb pseudopotential (${\mu^*}$) in
${{\rm C}_{60}}$ for a variety
of bare potentials. We find a large reduction in ${\mu^*}$ due to intraball
screening alone, leading to an interesting non-monotonic dependence of
${\mu^*}$ on the bare interaction strength.
 We find that ${\mu^*}$ is positive for physically
reasonable bare parameters, but small enough
to make the electron-phonon coupling a viable mechanism for
superconductivity in alkali-doped fullerides. We end with some open
problems.
\end{abstract}
The discovery of the icosahedrally symmetric molecule ${{\rm C}_{60}}$ in
1985\cite{c60}, and
the subsequent discovery of the many interesting electronic properties
of doped fullerene lattices has led to an explosion of research work on
these compounds. In this Letter we will be concerned with the effects of
Coulomb interactions on fullerene
superconductivity.

 ${{\rm C}_{60}}$ forms an FCC
lattice which can be doped with up to 6 alkali metal atoms per
${{\rm C}_{60}}$. The triply-doped
compounds $A_3{{\rm C}_{60}}$ (where $A=K,\ Rb,\ Cs$) are bad metals but
superconduct at relatively high temperatures ($\approx 20K$ for
$K_3{{\rm C}_{60}}$)\cite{c60sc,c60exrev,c60threv}. They also have a
very short coherence length of $\xi\approx
30\AA$, which is only about 2 lattice spacings of the FCC
lattice.  Experimental
evidence\cite{bethune,haddon,ecklund,mitch,prass,ramirez},
supported by numerous phonon
calculations\cite{lanoo,varma,schluter,jishi,chen,faulhaber,assaphon,manini},
indicates that some of the intraball phonon modes are very high in energy and
sufficiently strongly coupled to the electrons to account for these high
$T_c$\ s.  Most of the phonon calculations assume that the Coulomb
repulsion between the electrons comprising a pair are  small, as they
are in a usual metallic superconductor. (Effective Coulomb
interactions are typically parameterized by a dimensionless number
${\mu^*}=N(E_F)U^*$\cite{morel}, where $N(E_F)$ is the density of states at the
Fermi
Surface  (FS) and $U^*$ is an effective interaction matrix element between
electrons at the FS. For a metal ${\mu^*}\approx0.2$).
However, there are two significant differences in the case of
superconducting fullerides which make this  correspondence
uncertain.  Firstly, the coherence length, which is a measure of how close the
electrons in a pair are,  is an
order of magnitude shorter in  $A_3{{\rm C}_{60}}$ than  in metallic
superconductors.
Furthermore, in metallic superconductors the  effective Coulomb
interaction between the electrons comprising a pair is reduced
by the highly retarded nature of the phonon-mediated attraction, as
shown by Anderson and Morel in 1962\cite{morel}
\begin{equation}
{\mu^*}=\frac{\mu}{1+\mu\log(W/{\omega}_d)}
\end{equation}
where ${\omega}_d$ is the Debye frequency,  $W$ is the electronic
bandwidth, and $\mu$ measures the instantaneous Coulomb repulsion.
In metals ${\omega}_d$ is typically two orders of magnitude lower
than $W$ , which accounts for the small ${\mu^*}$. However, the electronic
bandwith of the conduction level of $A_3{{\rm C}_{60}}$ is about $W\approx
0.5eV$,
while the phonon modes go up to an energy of $0.2 eV$\cite{c60exrev}, which
makes
retardation highly ineffective in reducing ${\mu^*}$.

However, it has been proposed by Baskaran and Tosatti\cite{bask}, and
simultaneously by Chakravarty and Kivelson\cite{chak}, that a repulsive bare
interaction between the electrons on a single fullerene molecule
can result in an effective
attraction between electrons in the conduction level. Second-order
perturbation theory (PT) in $U/t$ in the tight-binding model on the
truncated icosahedral (TI) lattice with the on-site Hubbard interaction
$U$  shows  that ${\mu^*}$ becomes negative beyond  $U/t\approx
3$\cite{chak}.
Exact
diagonalizations  on small clusters\cite{fye,white} also demonstrate the
possibility of
${\mu^*}<0$ for intermediate $U/t$. However, increasing the range of the
interaction increases ${\mu^*}$ in second-order PT\cite{goff,assa}, and the
validity of
second-order PT is questionable in the region where ${\mu^*}<0$\cite{assa}.

A numerical RG method has been developed recently by one of us to
address these issues and get a reliable estimate of ${\mu^*}$\cite{taku}. This
is
based on the extention of Wilson's momentum shell
integration\cite{wils} to the
case of condensed fermion systems as developed by
Shankar\cite{shankar}.
The crucial
distinction between the RG as applied to critical phenomena in
statistical mechanics, and the RG as applied to condensed fermion
systems  is that in the latter case there are an
infinite number of flowing coupling constants which are relevant to the
low-energy behavior of the system (in infinite volume)\cite{shankar}.
To illustrate the method for  finite sizes,
let us consider the case of interacting electrons on ${{\rm C}_{60}}$. We
start with the Hamiltonian
\begin{equation}
H = -t\sum_{\langle ij,s\rangle}(c^\dagger_{is}c_{js}+h.c.)
+\sum_{i,j}V(i,j) n_i n_j
\label{H}
\end{equation}
where the first term is just the tight-binding approximation and the
second term represents a generic density-density interaction between the
electrons. (We have also performed calculations for two different hopping
matrix elements for the purely pentagonal versus the other
bonds\cite{c60threv}, but
since the results are almost identical to the ones for the above
hamiltonian, we report only the latter here).
  We solve the tight-binding problem and get 60 eigenstates,
each belonging to a representation of the icosahedral group $I_h$\cite{icos}.
 We can then go to the zero-temperature fermionic path integral with the
action
\begin{eqnarray}
S &=& \sum_{ks}\int_{-\infty}^{\infty}{\bar{\eta}_{ks}}{\cal Z}^{-1}(k)
(i\omega-\epsilon(k))\eta_{ks}(\omega) \nonumber \\
&-&\sum_{\{k_i\},ss'}
\int_{-\infty}^{\infty} {d\omega_1\cdots d\omega_4\over(2\pi)^3}
\delta(\omega_1+\omega_2-\omega_3-\omega_4)
V(\{k_i\}){\bar{\eta}}_{k_1s}
{\bar{\eta}}_{k_2s'}\eta_{k_3s'}\eta_{k_4s}
\label{S}
\end{eqnarray}
where $\eta,\ {\overline{\eta}}$ are Grassmann variables representing the
fermions, and $V_{ss'}(k_1,k_2,k_3,k_4)$ is the matrix element of the
bare interaction, and $\cal Z$ is the quasiparticle residue, or the wave
function renormalization. For the bare theory ${\cal Z}=1$ and the
energies are the tight-binding energies. This method produces results in
excellent agreement with exact diagonalizations for small
clusters\cite{taku}.

We now integrate out the variables corresponding to the high energy
levels (levels far from the conduction level) step by step. We go to
two-loop order in the interaction matrix elements to compute effective
values for  the energy, the
wave function renormalization, and the interaction. Each
integration generates effective interactions for the remaining
levels\cite{taku}.
We continue until only the conduction level remains, and obtain an
effective hamiltonian for the conduction level. In accordance with the
formalism for infinite fermion systems\cite{shankar},
we allow all the independent
couplings to flow separately. In order to reduce computation time we use
the full symmetries of $I_h$ to keep only symmetry-reduced matrix
elements, in analogy with the Wigner-Eckhart theorem.
This still leads to $\approx10^5$ flowing couplings.
We also  ignore the frequency dependence of the
effective interactions (they are irrelevant in the RG sense for
properties near the fermi surface\cite{shankar}).
Apart from keeping track of $\cal Z$, going to
two-loop order enables us to  generate an internal validity criterion.
As far as
the effective interactions in the conduction level are concerned, the
smallest energy denominators come from the excitations to levels close
to the conduction level, which will be integrated out during the final
steps of the RG. We compare the two-loop to the one-loop contribution to
the effective interactions at
the final RG step. When their ratio becomes  $\approx 1$ we know that all
loops will be important, and that truncating to two-loop is invalid.
As will be evident from our numerical results, this method is reliable
up to intermediate values of $U/t\approx5$, which fortunately includes
the physical regime for ${{\rm C}_{60}}$\cite{c60threv}.

We have considered two classes of models for the bare potential; the
extended Hubbard model with on-site and nearest neighbor interactions, a
screened Ohno potential. We also considered the problem of
only two additional electrons in
the $T_{1u}$ conduction level (as long as the three- and higher-body
interactions are negligible compared to the two-body interaction, these
results should hold for arbitrary numbers of electrons). The two
additional electrons can be in one of three channels $A_g$, $T_{1g}$,
$H_g$, which roughly correspond to total angular momentum $L=0,\ 1,\ 2$
respectively\cite{icos}. Let us now proceed to the results.

In Fig.\ \ref{fig1} we show the effective interaction $U^*/t$
 for the on-site Hubbard model as a function of the bare interaction
$U/t$.  From the non-monotonicity it is very evident that no naive
correspondence can be made with the Anderson-Morel formula, eq(1). For $U/t<5$
the triplet $T_{1g}$ pair state has the lowest energy. When considered
in the  context of a lattice problem, this will lead to itinerant
ferromagnetism. There is also a level crossing between the $A_g$ and the
$H_g$ pair states at $U/t\approx3.5$,
and for $U/t>5 $ the $A_g$ state is the ground
state. Fig.\ \ref{fig2} shows how far one can trust these results by plotting
the ratio of the third order to the second order at the final step of RG
as a function of $U/t$. We see that these results are reliable up to
$U/t\approx5-6$. For $U/t>6$ the $A_g$ channel shows negative $U^*$, which
was precisely the original counterintuitive claim that inspired this
work. However, it unfortunately happens in a region where we cannot
trust the result. Fig.\ \ref{fig3} compares various different approximations
for
$U^*/t$ for the $A_g$ channel,
from which it is clear that perturbation theory fails for
physically interesting $U/t\approx3-5$, and even one-loop RG differs
significantly from the two-loop result. Finally, the ladder
resummation\cite{gunnar},
which corresponds in our language to a one-loop RG with only the Cooper
channel, is also significantly different from the two-loop result.
Qualitatively similar results hold for the other two channels.

The next set of three figures shows the same information for the
screened Ohno potential\cite{goff} given by
\begin{equation}
V(i,j)= {\frac{U\exp-(r_{ij}/\lambda)}{\sqrt{1+r_{ij}^2/a^2}}}
\end{equation}
with the screening length ($\lambda$) set to $0.75$ of a $C-C$ bond
length($=a$) (this $\lambda$  is somewhat larger than
 the Thomas-Fermi screening length for $A_3{{\rm C}_{60}}$\cite{c60threv}). The
major differences from the Hubbard model are that $U^*$ is generally
higher,  and  that the region of $U^*<0$ has been pushed
to higher $U/t$, as was predicted on the basis of second-order PT
earlier\cite{goff,assa}. It is easy to see from Fig.\ \ref{fig5}
that the method is
trustworthy only to about $U/t\approx3-4$ now, especially for the
$T_{1g}$ and $H_g$ channels. Once again Fig.\ \ref{fig6} shows the
comparison between the different approximation methods. Adding a nearest
neighbor interaction to the on-site Hubbard model leads to results very
similar to Fig.\ \ref{fig4}.

A simple physical picture of the reduction of $U$ to  $U^*$ is that the
renormalized
conduction level electrons interact weakly with the core primarily via the
lowest $H_g$ and $T_{1g}$ collective modes\cite{plasmon} (the second is
actually a spin wave),
which act very much like phonons in producing an effective interaction
between the added electrons. While this picture produces qualitatively
correct results, {\it i.e.}, that the $A_g$ pair state has the highest
negative curvature, and that the $T_{1g}$ pair state always has positive
curvature, the  issue is complicated by nonzero overlaps between occupied
and unoccupied levels in the interacting theory. We hope to  present
 a full analysis in a future publication.

 To summarize, we have used a novel finite-size RG method to compute the
effective Coulomb interactions at the FS for a single ${{\rm C}_{60}}$
molecule. The method has an internal self-consistency check, and is
reliable in the physically interesting range of bare interaction
strengths, as opposed to other approximation schemes. The results show
that while the effective interactions are always repulsive for
physically relevant bare interaction strengths, there is a dramatic
reduction in strength of the repulsion compared to its bare value. It is
fascinating to speculate that for $U/t>6$ in the Hubbard model two
added electrons will attract each other on ${{\rm C}_{60}}$, although the
method
has reached the limit of its validity here. Clearly,
 this
renormalization cannot be parametrized by the Anderson-Morel formula.

A useful feature that we have not yet exploited is to reduce the size of
the problem, especially for strong couplings. One could run the RG as
long as the validity criterion allowed one to, and then exactly
diagonalize the much smaller effective problem.

The biggest deficiency of the above results is that since they
include lattice screening phenomenologically at best (via Thomas-Fermi
static screening\cite{screen}),
they cannot be  applied with much confidence to
superconducting $A_3{{\rm C}_{60}}$. We are currently working on a RG scheme
which
will include the effects of polarization and charge fluctuations of
neighboring fullerene molecules on $U^*$, which should go some way
towards addressing the lattice screening
problem\cite{lammert}.
In the absence of such a calculation one can believe that
lattice screening should only reduce $U^*$, and that the viability of
phonon mediated theories would be enhanced. In the Fig.\ \ref{fig7} we
present a model calculation of the coupled electron-phonon system, where
the effective interactions at the conduction level in the Hubbard model
have been taken into account. We have used the phonon coupling constants of
Ref\cite{varma}.  Despite the repulsive Coulomb
interactions, the phonons succeed in binding the pairs.

There are many directions in which this work could be taken. One could
calculate the exciton energies\cite{exciton} for neutral and charged ${{\rm
C}_{60}}$, one could
measure the spin structure of neutral ${{\rm C}_{60}}$, which has been the
subject
of some  controversy\cite{spin}, and one could hope to elucidate
metallic screening on the lattice.

\acknowledgements

It is a pleasure to thank Prof. R. Shankar and Prof. A. Auerbach for
stimulating discussions. This work was supported in part by NSF grant
number  DMR-9311949.

%\end{document}

\begin{figure}
\caption{The Coulomb pseudopotential $U^*/t$ for the on-site Hubbard model
as a function of bare coupling strength $U/t$ for the $A_g$, $T_{1g}$,
and $H_g$ channels.}
\label{fig1}
\end{figure}

\begin{figure}
\caption{The ratio of the two-loop to one-loop contributions at the
final RG step as a function of $U/t$ for the various channels for the
on-site Hubbard model.
Note that renormalized potential
matrix elements are used in the loop calculations so that the result is
reliable for much greater $U/t$ than perturbation theory.}
\label{fig2}
\end{figure}

\begin{figure}
\caption{A comparison of different approximation methods for $U^*/t$ in
the $A_g$ channel. All the methods differ appreciably from our two-loop
RG results long before the two-loop RG becomes unreliable.}
\label{fig3}
\end{figure}

\begin{figure}
\caption{The Coulomb pseudopotential $U^*/t$ for the Ohno potential with
screening length $\lambda=0.75 a\approx1\AA$
as a function of bare coupling strength $U/t$ for the $A_g$, $T_{1g}$,
and $H_g$ channels.}
\label{fig4}
\end{figure}

\begin{figure}
\caption{The ratio of the two-loop to one-loop contributions at the
final RG step as a function of $U/t$ for the various channels for the
Ohno potential. }
\label{fig5}
\end{figure}

\begin{figure}
\caption{The same comparison as fig. 2 for the Ohno potential. Once
again, all the methods differ appreciably from our two-loop
RG results long before the two-loop RG becomes unreliable.}
\label{fig6}
\end{figure}

\begin{figure}
\caption{ A model calculation of the binding energy of two electrons in
$eV$,
including the effective $U^*$ from the electronic calculation, and
including  up to two vibrons \protect\cite{manini}. We have
used $t=2\ eV$ and the electron-phonon couplings
from Ref. \protect\cite{varma}. It is clear that even in the Ohno case the
phonons succeed in binding the electrons.}
\label{fig7}
\end{figure}

\end{document}